\documentclass[aps,prd,preprint,superscriptaddress,nofootinbib]{revtex4-1}
\pdfoutput=1
\usepackage{graphicx}  
\usepackage{subfigure}
\usepackage{latexsym}
\usepackage{slashed}
\usepackage{dcolumn}   
\usepackage{bm}        
\usepackage{amsmath}
\usepackage{amssymb}   
\usepackage{longtable}
\usepackage[section]{placeins}
\usepackage{float}

\newcommand{\uva}{\affiliation{Department of Physics, University of Virginia, Charlottesville, VA 22904-4714, USA}}

\begin{document}

\title{Examining the Viability of Phantom Dark Energy}

\author{Kevin J. Ludwick}
\email{kludwick@virginia.edu} \uva

\begin{abstract}

In the standard cosmological framework of the 0th-order FLRW metric and the use of perfect fluids in the stress-energy tensor, 
dark energy with an equation-of-state parameter $w < -1$ (known as phantom dark energy)
implies negative kinetic energy and vacuum instability when modeled as a scalar field.  
However, the accepted values for present-day $w$ from Planck and WMAP9 include a significant range of values less than $-1$.  We find that it is not as obvious as one might think that phantom 
dark energy has negative kinetic energy categorically.  Analogously, we find that field models of quintessence dark energy ($w_\phi>-1$) do not necessarily 
have positive kinetic energy categorically.  Staying within the confines of observational constraints 
and general relativity, for which there is good experimental validation, we consider 
a few reasonable departures from the standard 0th-order framework in an attempt to see if negative kinetic energy can be avoided 
in these settings despite an apparent $w<-1$.  
We consider a more accurate description of the universe through the perturbing of the isotropic and homogeneous FLRW metric and the components 
of the stress-energy tensor,  and we
consider dynamic $w$ and primordial 
isocurvature and adiabatic perturbations.  We find that phantom dark energy does not necessarily have negative kinetic energy 
for all relevant length scales at all times, and we also find that, by the same token, quintessence dark energy does not necessarily have positive 
kinetic energy for all relevant length scales at all times.


\end{abstract}

\maketitle

\renewcommand{\thepage}{\arabic{page}}
\setcounter{page}{1}


\begin{center}
{\bf Introduction}
\end{center}

A recent milestone in observational cosmology happened when the High-z Supernova Search Team in 1998 \cite{Riess} and the Supernova Cosmology Project in 1999 \cite{SCP} published observations of the emission spectra of Type Ia supernovae indicating that the universe's rate of outward expansion is increasing.  
Galaxy surveys and the late-time integrated Sachs-Wolfe effect also give evidence for the universe's acceleration.  Thus, "dark energy" was proposed 
as the pervasive energy in the universe necessary to produce the outward force that causes this acceleration, which has been observationally tested and vetted 
since its discovery.  The 2011 Nobel Prize in Physics was awarded to Schmidt, Riess, and Perlmutter for their pioneering work leading to the discovery of dark energy.  
The present-day equation-of-state parameter $w$ from the equation of state most frequently tested by cosmological probes, $p=w \rho$ with constant $w$, assuming a 
flat universe and a perfect fluid representing dark energy, has been constrained by Planck in early 2015 to be $w=-1.006 \pm 0.045$ \cite{Planck}, and Planck's 2013 
value was $w=-1.13^{+0.13}_{-0.10}$ \cite{Planck2013}.  The value from the Nine-Year Wilkinson Microwave Anisotropy Probe (WMAP9), combining data from 
WMAP, the cosmic microwave background (CMB), baryonic acoustic oscillations (BAO), supernova measurements, and $H_0$ measurements, is 
$w = -1.084 \pm 0.063$ \cite{WMAP9}.  From these reported values, the prospect of $w<-1$ is clearly a distinct possibility, and under other assumptions (such as a spatially curved universe), 
the window reported for $w$ does not always include the value for the cosmological constant (CC) model, $w=-1$.  

However, dark energy modeled as a perfect fluid with $w < -1$ leads to a field theory with negative kinetic energy (a ghost field theory), which implies vacuum instability.  Either 
the phantom ghost has positive density and violates unitarity, rendering it unphysical, or unitarity is satisfied and the density is negative, which leads to vacuum 
instability \cite{Cline}.  This phantom dark energy with a wrong-sign kinetic term described as an 
effective field theory may be able to make this instability unobservable, but not without great 
difficulty and perhaps sacrifice of well-accepted physical principles \cite{Cline, CHT}.  

One deduces the ghost nature of phantom dark energy from $w<-1$ within the standard cosmological framework of the 0th-order Friedmann-Lema\^{i}tre-Robertson-Walker 
(FLRW) metric with the use of perfect fluids 
in the stress-energy tensor, but the condition for negative kinetic energy is different for different frameworks.  In this work, given that our universe is not 
perfectly isotropic and homogeneous, we examine the possibility for positive kinetic energy with $w<-1$ in light of first-order perturbations to 
the FLRW metric
 and the components of the stress-energy tensor.  
We first consider constant $w$ and then dynamic $w$.  
 We also consider primordial 
 isocurvature along with adiabatic perturbations.  We then consider the possibility of negative kinetic energy for $w>-1$ via the inclusion of cosmological perturbations.  
Although there are many alternative frameworks one may study (inclusion of spatial curvature, dynamic sound speed $c_s$, vector field dark energy, models with 
coupled dark energy and dark matter, modified gravity, different metrics, 
dark energy as an imperfect fluid, quantum corrections), we focus on the aforementioned manifestations of dark energy in this paper and leave these others to future work.  
There is abundant experimental verification of general relativity, 
so we study the condition on the kinetic energy of phantom dark energy within the confines of general relativity and 
observational constraints from cosmological probes.  

\bigskip
\begin{center}
{\bf Phantom Dark Energy}
\end{center}

Consider the Einstein-Hilbert action for general relativity with a complex scalar field ($c=1$):
\begin{equation}
\label{action}
S = \int d^4 x \sqrt{-g} \left[ \frac{R}{16 \pi G} - \frac{1}{2} g^{\mu \nu} \nabla_\mu \phi^* \nabla_\nu \phi - V(|\phi|) \right] + S_m,
\end{equation}
where the first term is the usual contribution to the Einstein tensor, the second and third terms are the contribution to the scalar field dark energy, and $S_m$ is the action 
for the rest of the components of the stress-energy tensor $T_{\mu \nu}$.  Minimizing the action leads to Einstein's equation, 
\begin{equation}
\label{Einstein}
R_{\mu \nu} - \frac{1}{2} R g_{\mu \nu} = 8\pi G(T_{\mu \nu}[\phi]+T_{\mu \nu}[m]),
\end{equation}
where $T_{\mu \nu}[\phi] = -2 \frac{\delta \mathcal{L}_\phi}{\delta g^{\mu \nu}}+ g_{\mu \nu} \mathcal{L}_\phi$.

Assuming dark energy is spatially homogeneous as a perfect fluid, the density $\rho_\phi$ and pressure $P_\phi$ for the scalar field are  
\begin{equation}
\label{scalarfluid}
\rho_{\phi} = \frac{\dot{|\phi|^2}}{2 a^2}+ V(|\phi|), \quad P_{\phi} = \frac{\dot{|\phi|^2}}{2 a^2} - V(|\phi|).
\end{equation}
We used the flat FLRW metric
\begin{equation}
\label{FLRW}
ds^2= a^2(\tau) \left[- d\tau^2 + dx^i dx_i \right],
\end{equation}
and $\cdot$ represents differentiation with respect to $\tau$.  

The kinetic energy for the scalar field from the Lagrangian density $\mathcal{L}_\phi$ is $ - \frac{1}{2} g^{\mu \nu} \nabla_\mu \phi^* \nabla_\nu \phi = \frac{\dot{|\phi|^2}}{2 a^2}$.  
The equation-of-state parameter $w=\frac{P}{\rho}$ for dark energy is 
\begin{equation}
\label{w}
w_\phi=\frac{\frac{\dot{|\phi|^2}}{2 a^2} - V(|\phi|)}{\frac{\dot{|\phi|^2}}{2 a^2} + V(|\phi|)},
\end{equation}
and one can see that $w_\phi<-1$ and the physically reasonable condition $\rho_\phi \geq 0$ imply $\rho_\phi+P_\phi=\frac{\dot{|\phi|^2}}{a^2} = 2 ~\mathrm{KE}_\phi < 0$, which 
cannot be true for a complex or real scalar field.  So in order for $w_\phi<-1$ to be consistent with the positivity of $\frac{\dot{|\phi|^2}}{a^2}$, the 
usual procedure is to flip the sign of the kinetic term in the Lagrangian density so that the dark energy density is $-\frac{\dot{|\phi|^2}}{2 a^2} + V(|\phi|)$ and the pressure is 
$-\frac{\dot{|\phi|^2}}{2 a^2} - V(|\phi|)$.  This is what is done in k-essence dark energy, and as we have mentioned, the negative kinetic term results in vacuum instability and difficulties in framing a viable effective field theory.  We do not use this framework.  
The positivity of the kinetic energy for the usual sign of the kinetic term, $2 ~\mathrm{KE}_\phi \geq 0$, 
is equivalent to $w_\phi \geq -1$, which is 
the Null Energy Condition ($\rho_\phi+P_\phi \geq 0$) for a single perfect fluid with positive density\footnote{Interestingly, the Null Energy Condition (NEC) is not 
violated for all $w_\phi<-1$ at present time since it requires $\sum_i(\rho_i + P_i) \geq 0$.  If one includes both dark energy and matter components at the present 
time, $w_\phi$ can be as low as $-1-\Omega_{m0}/\Omega_{DE0} = -1.46$ and still not violate the NEC.  However, dark energy would still have a negative kinetic term.  We do not frame our discussion 
primarily in terms of the classical energy conditions of general relativity; we are merely concerned with the sign of the kinetic energy.}.

In order to avoid the unwanted pathologies of phantom fields while still maintaining $w_\phi<-1$, perhaps the kinetic energy can be kept positive when examined under 
a perturbative departure from isotropy and homogeneity.  In a universe that we know is not perfectly isotropic and homogeneous, perhaps describing phantom dark energy 
as a field theory in such a 0th-order framework is neither appropriate nor physically accurate.  

If we perturb the field $\phi(\tau) \rightarrow \phi(\tau) + \delta \phi(\vec{x},\tau)$, the density and pressure are also perturbed 
($\rho_\phi(\tau) \rightarrow \rho_\phi(\tau) + \delta \rho_\phi(\vec{x},\tau)$, $P_\phi(\tau) \rightarrow P_\phi(\tau) + \delta P_\phi(\vec{x},\tau)$), and we have 
\begin{equation}
\label{perturbw}
w_{\mathrm{eff}} \equiv \frac{P_\phi + \delta P_\phi}{\rho_\phi + \delta \rho_\phi} = \frac{\frac{1}{2a^2}(\dot{|\phi|^2} + \dot{\phi^*} \dot{\delta \phi} + \dot{\phi} \dot{\delta \phi^*})- (V + V' \delta \phi)}{\frac{1}{2a^2}(\dot{|\phi|^2} + \dot{\phi^*} \dot{\delta \phi} + \dot{\phi} \dot{\delta \phi^*}) + (V +V' \delta \phi)},
\end{equation}
where $'$ denotes differentiation with respect to $\phi$ and $\mathrm{KE}_\mathrm{eff} = \frac{1}{2a^2}(\dot{|\phi|^2} + \dot{\phi^*} \dot{\delta \phi} + \dot{\phi} \dot{\delta \phi^*})$.  
In analogy with the 0th-order case, the positivity of the kinetic energy, 
$2 ~\mathrm{KE}_\mathrm{eff} \geq 0$, is equivalent to $w_{\mathrm{eff}} \geq -1$, which is 
the Null Energy Condition ($\rho_\phi + \delta \rho_\phi + P_\phi + \delta P_\phi \geq 0$) for a single perfect fluid with positive density.  

However, in the perturbative approach, the equations hold separately for 0th and 1st orders, and the 1st-order perturbation parameters are evaluated assuming the 0th-order 
equations.  For a fluid model of phantom dark energy, this approach is completely consistent, but it is not for a phantom scalar field because of the incompatibility at the 0th order 
we saw earlier.  Instead, we can define a scalar field theory for phantom dark energy in the background of the FLRW metric {\it only} when it is with 1st-order perturbations.  In that case, we define a field $\Phi$ valid only at 1st order by 
\begin{align}
\rho_\Phi(\vec{x},\tau) \equiv \rho_\phi(\tau) + \delta \rho_\phi(\vec{x},\tau) = \frac{|\dot{\Phi}|^2}{2a^2} -k^2 \frac{|\Phi|^2}{2}+V(|\Phi|), \label{rhoPhi} \\
P_\Phi(\vec{x},\tau) \equiv P_\phi(\tau) + \delta P_\phi(\vec{x},\tau) = \frac{|\dot{\Phi}|^2}{2a^2} -k^2 \frac{|\Phi|^2}{2}-V(|\Phi|), \label{PPhi} \\
2 \mathrm{KE}_\Phi = \rho_\Phi + P_\Phi = \frac{|\dot{\Phi}|^2}{a^2} -k^2 |\Phi|^2, \label{KEPhi}
\end{align}
where the term proportional to $k^2$ is present for a field $\Phi(\vec{x},\tau)$ that is not spatially homogeneous.  So for an apparent value of 
$w_\phi < -1$ as measured by observational probes, it may be the case that $w_\Phi \equiv P_\Phi/\rho_\Phi \geq -1$ and $\mathrm{KE}_\Phi \geq 0$, indicative of 
a viable scalar (real or complex) field theory for phantom dark energy\footnote{Note that the condition
 for positive KE, $\rho_\phi + \delta \rho_\phi + P_\phi + \delta P_\phi \geq 0$, has terms of zeroth and first order in the same inequality even though first order perturbations 
 are small compared to zeroth order terms.  However, this inequality is especially useful in studying the sign of the kinetic energy when the first order terms make a 
 non-negligible contribution to the inequality, namely when the zeroth order terms are collectively small.  This is the case when $|\rho_\phi + P_\phi| \ll 1$, or $|w_\phi +1| \ll 1$.}.  


\bigskip
\begin{center}
{\bf Cosmological Perturbation Theory}
\end{center}

The Friedmann equations resulting from solving Einstein's equation for the flat FLRW metric (0th order), Eq. (\ref{FLRW}), are 
\begin{align}
\mathcal{H}^2~ &= ~\frac{8 \pi G}{3} a^2 \rho, \label{Feq1} \\
\dot{\mathcal{H}}~ &= ~ - \frac{4 \pi G}{3} a^2 (\rho+ 3P), \label{Feq2} 
\end{align}
where $\mathcal{H} \equiv \frac{\dot{a}}{a}$ and $\rho$ and $P$ represent the sum of the density and pressure components respectively.  
These equations lead to the evolution equation for 
each density component:
\begin{equation}
\label{densityevo}
\dot{\rho} = -3 \mathcal{H} (\rho+ P).
\end{equation}

Throughout this work, we consider scalar perturbations in the synchronous gauge, and we use the notation of \cite{BertMa} in the following.  
The perturbed metric is 
\begin{equation}
\label{perturbFLRW}
ds^2 = a^2(\tau) \left[ -d\tau^2 + (\delta_{ij}+ h_{ij}) dx^i dx^j \right],
\end{equation}
and the scalar mode of $h_{ij}$ is written in $k$-space as 
\begin{equation}
\label{hij}
h_{ij}(\vec{x},\tau) = \int d^3k e^{i \vec{k} \cdot \vec{x}} \biggl\{\hat{k_i} \hat{k_j} h(\vec{k},\tau)+(\hat{k_i} \hat{k_j} -\frac{\delta_{ij}}{3}) 6 \eta(\vec{k},\tau) \biggr\}.
\end{equation}
The equations resulting from solving the perturbed Einstein equation in 
Fourier space to first order are
\begin{subequations}
\label{perturb}
\begin{align}
k^2 \eta - \frac{1}{2} \mathcal{H} \dot{h}~ &= ~ 4 \pi G a^2 \delta T^0_0, \label{perturb1} \\ 
k^2 \dot{\eta} ~ &= ~ 4 \pi G a^2 (\rho+P) \theta, \label{perturb2} \\
\ddot{h} + 2 \mathcal{H} \dot{h} - 2 k^2 \eta ~ &= ~ -8 \pi G a^2 \delta T^i_i, \label{perturb3} \\
\ddot{h} + 6 \ddot{\eta} +2 \mathcal{H} (\dot{h} + 6 \dot{\eta}) - 2k^2 \eta ~ &=~ -24 \pi G a^2 (\rho+P) \sigma, \label{perturb4}
\end{align}
\end{subequations}
where  $\theta$ is the divergence of the fluid velocity $v_i$, $(\rho + P) \sigma \equiv - (\hat{k}_i \hat{k}_j - \frac{1}{3} \delta_{ij}) \Sigma^i_{~j}$ where 
$\Sigma^i_{~j}$ is the anisotropic shear perturbation, and $h$ and $\eta$ are the scalar modes of the metric perturbation.  The stress-energy 
tensor is given by 
\begin{align}
T^0_{~0} ~ &= ~ -(\rho+\delta \rho),  \nonumber \\
T^0_{~i} ~ &=~ (\rho+P)v_i,  \nonumber \\
T^i_{~j} ~&=~ (\rho+ \delta P)\delta^i_{~j} + \Sigma^i_{~j}, \quad \Sigma^i_{~i} = 0. \label{stressenergy}
\end{align}

The conservation of energy-momentum, $T^{\mu \nu}_{~~;\mu} = 0$, gives (using $\delta \equiv \delta \rho/\rho$)
\begin{subequations}
\label{cons}
\begin{align}
\dot{\delta} ~&= ~ -(1+w) \left(\theta + \frac{\dot{h}}{2} \right)- 3 \mathcal{H} \left( \frac{\delta P}{\delta \rho} -w \right) \delta, \label{perturb5} \\
\dot{\theta} ~&= ~ -\mathcal{H}(1-3w)\theta - \frac{\dot{w}}{1+w} \theta + \frac{\delta P/\delta \rho}{1+w} k^2 \delta - k^2 \sigma. \label{perturb6}
\end{align}
\end{subequations}
Eq. (\ref{cons}) is valid when considering each fluid component or the total fluid, but Eq. (\ref{perturb}) is valid only 
for the total fluid.  The anisotropic shear stress is $0$ ($\sigma_\phi = 0$) throughout, and in what follows, we use $c=G=1$.  As is typically done to define the coordinates 
of the synchronous gauge, we choose $\theta_c=0$, for cold dark matter.

$\delta P/\rho$ for a given fluid component
is in general given by  
\begin{equation}
\label{deltap}
\frac{\delta P}{\rho} = c_s^2 \delta + (c_s^2-c_a^2) 3 \mathcal{H}(1+w) \frac{\theta}{k^2},
\end{equation}
where $c_s$ is the fluid's sound speed and $c_a^2 \equiv \dot{P}/\dot{\rho} = w + \dot{w} \rho/\dot{\rho}$ is defined as the square of the fluid's 
adiabatic sound speed \cite{0804.0232}.  For a barotropic fluid, $c_s^2 = c_a^2$, and $c_a^2 = w$ for constant $w$.  Even though dark energy can have a barotropic 
equation of state, 
treating it like an adiabatic fluid (for which Eq. (\ref{deltap}) reduces to $\delta P = c_s^2 \delta \rho$) would imply imaginary sound speed and instabilities in dark energy, 
so we use this general relation between $\delta P$ and $\delta \rho$.  

\bigskip

\begin{center}
{\bf Condition for Positive Kinetic Term}
\end{center}

We want to determine if $\rho_\phi + \delta \rho_\phi + P_\phi + \delta P_\phi \geq 0$ ($w_{\mathrm{eff}} \geq -1$) using observational constraints.  For positive density, this inequality is 
equivalent to 
\begin{equation}
\label{condition}
1+  \delta_\phi + w_\phi + \delta P_\phi/\rho_\phi \geq 0.
\end{equation}
Using Eqs. (\ref{deltap}) and (\ref{perturb6}) to put $\delta_\phi$ in terms of $\theta_\phi$ and $\dot{\theta}_\phi$, this inequality becomes
\begin{equation}
\label{condition2}
(1+w_\phi) \left( 1+ \frac{\mathcal{H}}{k} \left[ \frac{1+c_{s\phi}^2}{c_{s\phi}^2} \frac{dV_\phi}{da} a + V_\phi \left( \left\{ \frac{(1+c_{s\phi}^2)(1-3c_{s\phi}^2)}{c_{s\phi}^2} + 3c_{s\phi}^2-3w_\phi \right\} + \frac{dw_\phi}{da} \frac{a}{1+w_\phi} \right) \right] \right) \geq 0,
\end{equation}
where $V \equiv \theta/k$.  

For $w_\phi \leq -1$, Inequality (\ref{condition2}) is equivalent to 
\begin{equation}
\label{condition3}
\frac{\mathcal{H}}{k} \left[ \frac{1+c_{s\phi}^2}{c_{s\phi}^2} \frac{dV_\phi}{da} a + V_\phi \left( \left\{ \frac{(1+c_{s\phi}^2)(1-3c_{s\phi}^2)}{c_{s\phi}^2} + 3c_{s\phi}^2-3w_\phi \right\} + \frac{dw_\phi}{da} \frac{a}{1+w_\phi} \right) \right] \leq -1.
\end{equation}
For the case of a scalar field \cite{0804.0232}, 
\begin{equation}
\label{cs2}
c_{s\phi}^2 \equiv \frac{\delta P_\phi}{\delta \rho_\phi} \bigg|_{\mathrm{rest ~frame}} = 1.
\end{equation}
Inequality (\ref{condition3}) then reduces to
\begin{equation}
\label{conditionscalar}
\mu \equiv \frac{\mathcal{H}}{k} \left[ 2 \frac{dV_\phi}{da} a + V_\phi \left( \left\{ -1-3w_\phi \right\} + \frac{dw_\phi}{da} \frac{a}{1+w_\phi} \right) \right] \leq -1.
\end{equation}
Even with the constraints $|V_\phi| < 1$ and $|\dot{V_\phi}| = | a \mathcal{H}\frac{dV_\phi}{da}| <1$, it is mathematically possible for Inequality 
(\ref{conditionscalar}) to be satisfied, but we must solve for 
$V_\phi$ from Eqs. (\ref{perturb}) and (\ref{cons}) to 
determine for certain.  

\bigskip

\begin{center}
{\bf Dark Energy with Constant $w$}
\end{center}

For dark energy with constant $w$, Eqs. (\ref{perturb}) and (\ref{cons}) can be solved explicitly for different eras dominated by one fluid component.  Assuming one 
fluid component with constant $w$ during an era dominated by that fluid component, $\mathcal{H}$ is given by 
\begin{equation}
\mathcal{H} = \frac{2}{(3w +1) \tau}.
\end{equation}
For $w < -1/3$, $\tau \in (- \infty,0)$ for $a \in (0, \infty)$, whereas $\tau$ is positive for $w>-1/3$.  For working between different eras, we find it more convenient 
to work in terms of $a$ rather than $\tau$.  $\tau$ can be related to $a$ via Eq. (\ref{Feq1}).  

We know that the universe is homogeneous and isotropic only 
 on large scales; on smaller scales, the universe obviously is not described by a FLRW metric, even one with small 1st-order perturbations.  And since we are only concerned with 
 dark energy, which we know to act only on large scales at which its repulsive force is detectable relative to the attractive force of gravity that keeps smaller structures bound 
 together, we do not consider large $k$.  As one might expect, the assumption of 1st-order perturbations ($|\delta| \ll 1$, $|\delta P/P| \ll 1$, $|V_\phi|  \ll 1$, and their derivatives 
 with respect to $\tau$) 
 is not always satisfied for large $k$.  
 The spectra of Type Ia supernovae that have enabled the detection of universal acceleration span in redshift from $z \approx 0.3$ to $z \approx 2$ \cite{SCP2}, 
which corresponds to spectral shifting over a range of 
$10^{-4} ~\mathrm{Mpc}^{-1} \lesssim k \lesssim  8 \times 10^{-4} ~ \mathrm{Mpc}^{-1}$.  
The late-time Sachs-Wolfe effect, 
which is evidence for the effect of dark energy on superclusters and supervoids in the CMB, also occurs on similarly large scales.  
An acceptable theory of dark energy must at least be valid for a relevant range of large scales, at least for 
$10^{-4} ~\mathrm{Mpc}^{-1} \lesssim k \lesssim  8 \times 10^{-4} ~ \mathrm{Mpc}^{-1}$, and this is the range of $k$ we consider in the following analysis.  

\begin{center}
{\it Radiation Domination}
\end{center}

During radiation domination ($\mathcal{H} = 1/\tau$ and $\tau = a (\frac{3}{\rho_{r0} 8\pi })^{1/2}$), $V_\phi^{rad}$ with constant $w_\phi \neq -1$ and sound 
speed $c_{s \phi}$ is obtained from Eq. (\ref{cons}) and from the non-gauge 
mode for $h$ for $k<< \mathcal{H}$ during the radiation era \cite{BertMa}, 
$h=A (k \tau)^2$.  
We set the constants of integration for the decaying modes of the full solution for $V_\phi^{rad}$ to $0$, and this gives a real solution for $V_\phi^{rad}$.  
For $k \ll \mathcal{H}$, $V_\phi^{rad}$ is 
\begin{equation}
\label{Vrad2}
V_\phi^{rad} \approx -\frac{3^{3/2} a^3 A c_{s \phi}^2 k^3 }{2^{11/2} (\pi \rho_{r0})^{3/2} (4+3 c_{s \phi}^2-6 w_\phi)}, 
\end{equation}
where $A$, which is in general a function of $k$, is a constant of integration with respect to $\tau$ (from the expression for $h$) 
and $\rho_{r0} = \frac{3 H_0^2}{8 \pi} \Omega_{r0}$ is the 
present-day density of radiation, which we define 
to be at $a=1$.  We use $H_0=67.3$ km/s/Mpc and $\Omega_{r0} = 9.24 \times 10^{-5}$, both of which are consistent with Planck's reported values \cite{Planck}.  

Since the comoving curvature perturbation $\mathcal{R}$
equals the square root of the scalar power spectrum
$\pm \sqrt{\mathcal{P}_{\mathcal{R}}} = \pm \sqrt{A_s \biggl(\frac{k}{k_{\star}}\biggr)^{n_s-1}}$ for adiabatic perturbations, and since 
\begin{equation}
\label{comoving}
\mathcal{R} = -\eta + \frac{\mathcal{H}}{k^2}[\dot{h} + 6 \dot{\eta}] + \mathcal{H} \frac{V}{k} 
\end{equation}
in the synchronous gauge, we can use Planck's constraint on $A_s$ and $n_s$ 
and expressions for $\eta$ and $h$ to get a value for $A$ in $V_\phi^{rad}$.  

On distance scales for which linear perturbation theory applies (roughly $k > 0.1~ {\mathrm{Mpc}}^{-1}$), we are agnostic concerning the sign of $\mathcal{R}$ 
since it appears only in squared form in all observations, including the CMB angular power spectrum and the matter power spectrum.  The sign of $A$, which appears 
in all the expressions for super-horizon perturbations in the early universe, determines the sign of density and velocity perturbations, and $A$ is determined by 
the sign of $\mathcal{R}$.  
The Press-Schecter formalism is a good approximate description of the formation of galaxies and clusters, and it assumes positive matter density perturbations.  However, 
it operates on smaller scales for which linear perturbation theory breaks down and higher order terms may dominate.  So we will deal with both signs in our analysis.

From \cite{BertMa}, for adiabatic initial conditions and $k \ll \mathcal{H}$ during radiation domination,
\begin{equation}
\label{metricperturbs}
\eta = 2 A - \frac{5+4 R_{\nu}}{6 (15+4R_{\nu})} A (k \tau)^2, ~~ h = A (k \tau)^2,
\end{equation}
and 
\begin{equation}
\label{initialv}
V = (1-R_{\nu})V_{\gamma}+ R_{\nu} V_{\nu} = -\frac{1}{18} A (k \tau)^3 \biggl(1- R_{\nu}+R_{\nu} \frac{23+4 R_{\nu}}{15+4R_{\nu}}\biggr),
\end{equation} 
where $R_{\nu} \equiv \frac{\rho_{\nu}}{\rho_{\gamma}+\rho_{\nu}}$ is the contribution of neutrino density out of the total radiation density from photons and neutrinos.  Using 
our aforementioned value of $\rho_{r0}$, $\rho_{\nu}/\rho_{\gamma}=(7 N_{\nu}/8)(4/11)^{4/3}$, and Planck's value of $N_{\nu}=3.046$, we get $R_{\nu} = 0.409$.  
Combining Eqs. (\ref{comoving}), (\ref{metricperturbs}), and (\ref{initialv}) gives 
\begin{equation}
\label{IC}
A(k) = \frac{\mp \sqrt{\mathcal{P}_{\mathcal{R}}} (15+4R_{\nu})}{2(5+4R_{\nu})}.
\end{equation}
We see that $\mathcal{R}$ is constant with respect to $\tau$ on super-horizon scales.  We use $\ln(10^{10} A_s) = 3.064 \pm 0.023$, $n_s = 0.9667 \pm 0.0040$, and a pivot 
scale of $k_{\star} = 0.05~ \mathrm{Mpc}^{-1}$ from Planck's 2015 results \cite{Planck}.  For most values of relevant $k$ we consider, $A \sim \mp 10^{-5}$.

Using Eq. (\ref{Vrad2}), Ineq. (\ref{conditionscalar}) becomes 
\begin{equation}
\label{Vrad2ineq}
\mu = \frac{3 A k^2a^2 \left(c_{s \phi}^2 (3 w_\phi -1)-4\right)}{16 \pi  \rho_{r0}\left(3 c_{s _\phi}^2-6 w_\phi +4\right)} \leq -1.
\end{equation}
We see that for $w<-1$, the inequality can only be satisfied if $A>0$.  

In Ineq. (\ref{conditionscalar}), $A$ is a prefactor of the entire lefthand side, so the magnitude of $A$ largely controls whether the inequality is met.  
Without constraints on $A$, we can find a scenario in which 
Ineq. (\ref{Vrad2ineq}) is satisfied for some ranges of parameters, and models of dark energy with $w_\phi<-1$ do not obviously have a negative kinetic term.  
For example, for $k=10^{-4} ~\mathrm{Mpc}^{-1}$, $w_\phi=-1.1$, and $A = 44445$, the inequality is satisfied 
for $a > 1.9 \times 10^{-4}$, and $|V_\phi^{rad}|,~ |\dot{V}_\phi^{rad}|= |a \mathcal{H}~ V^{rad}_{\phi}~ '(a)|,~ |\delta_\phi^{rad}|, ~ |\delta P_{\phi}/P_{\phi}|<1$ throughout.  
However, when we 
constrain $A$ for adiabatic initial conditions, we get that $|A|$ is too small for 
Ineq. (\ref{Vrad2ineq}) to be satisfied.  
$\mu$ is made most negative for 
biggest $w_\phi$ ($w_\phi \rightarrow -1$), maximal $a$ ($a=a_{eq}=1/3043$, where $a_{eq}$ is when the density of radiation and matter are equal), and maximal $k$ for which $k \ll \mathcal{H}$ is still accurate 
($k \approx \mathcal{H}(a_{eq}) = 0.0066~ \mathrm{Mpc}^{-1}$).  
However, when we use these parameters 
with our value of $A(k = \mathcal{H}(a_{eq}))$ from Eq. (\ref{IC}), $\mu \sim 
-3 \times 10^{-6}$, and Ineq. (\ref{Vrad2ineq}) 
is clearly not satisfied.  

The assumption of isocurvature perturbations changes the contribution from perturbations.  Isocurvature perturbations are fairly tightly constrained by 
observations.  The maximally allowed contribution from isocurvature modes $\alpha_{\mathrm{non-adi}}$ from observational constraints from Planck \cite{Plancki} are
\begin{align}
&\mathrm{cold~ dark~ matter~ density~ mode:} ~ (-1.5\%,~1.9\%), \nonumber \\
&\mathrm{neutrino ~density~ mode:} ~ (-4.0\%,~1.4\%), \nonumber \\
&\mathrm{neutrino ~velocity ~mode:} ~ (-2.3\%,~2.4\%). \nonumber 
\end{align}
Using the maximal amount of isocurvature contribution in each case to maximize the magnitude of $\mu$ has a very negligible 
effect on perturbations' values.  

The perturbations in the radiation era for different isocurvature modes are given in several places in the literature.  For the synchronous gauge, $h$, 
$\delta_c$, and $\delta_\gamma$ in the 
cold dark matter density isocurvature mode (CDI) are initialized by \cite{isoconsts}
\begin{align}
\label{CDI}
&h = A_{CDI} \frac{\Omega_{c0}}{\sqrt{\Omega_{r0}}} H_0 \tau - \frac{3}{8} A_{CDI} \frac{\Omega_{c0} \Omega_{m0}}{\Omega_{r0}} H_0^2 \tau^2 \nonumber \\
&\delta_c = A_{CDI} - \frac{h}{2} \nonumber \\
&\delta_\gamma = - \frac{2}{3} h, 
\end{align}
where $\Omega_{c0}$ and $\Omega_{m0}$ are the present-day cold dark matter and total matter density contribution respectively.  
We can then obtain the initial condition for $V_{\phi}^{rad}$ for $k \ll \mathcal{H}$ for the CDI mode using Eq. (\ref{cons}) with $h$ from Eq. (\ref{CDI}).  
Using a best fit from Planck's TT spectrum and low-$l$ 
polarization modes \cite{Plancki}, the magnitude of the primordial power spectrum at a scale of $k_1=0.002~ \mathrm{Mpc}^{-1}$ is 
$\mathcal{P}^1_{\mathcal{I} \mathcal{I}} = 1.4 \times 10^{-11}$ for the 
CDI mode and $\mathcal{P}^1_{\mathcal{R} \mathcal{R}} = 2.4 \times 10^{-9}$ for the adiabatic mode, and at $k_2 = 0.100 ~\mathrm{Mpc}^{-1}$,  
$\mathcal{P}^2_{\mathcal{I} \mathcal{I}} = 4.7 \times 10^{-13}$ 
and $\mathcal{P}^2_{\mathcal{R} \mathcal{R}} = 2.1 \times 10^{-9}$.  They use
\begin{equation}
\label{PlanckPS}
\mathcal{P}_{ab}(k) = \mathrm{Exp}\biggl[\biggl(\frac{\ln k - \ln k_2}{\ln k_1-\ln k_2}\biggr) \ln (\mathcal{P}^1_{ab})+
\biggl(\frac{\ln k - \ln k_1}{\ln k_2-\ln k_1}\biggr) \ln (\mathcal{P}^2_{ab}) \biggr]
\end{equation} 
as their parameterization for the primordial power spectrum, and $a,~b = \mathcal{R},~\mathcal{I}$.  
We can specify the constant $A_{CDI}$ using Eq. (\ref{CDI}), Eq. (\ref{PlanckPS}), and the relationship between the entropy perturbation and the 
isocurvature primordial power spectrum, 
\begin{equation}
\mathcal{S}_\alpha \equiv \frac{\delta_\alpha}{1+w_\alpha} - \frac{\delta_\gamma}{1+w_\gamma} = \pm \sqrt{\mathcal{P}_{\mathcal{I} \mathcal{I}}}
\label{entropyperturb}
\end{equation}
for the isocurvature mode denoted by $\alpha$.  After specifying $A_{CDI}$ from data, we can use $h$ from the CDI mode to solve for $V_\phi^{rad}$ in the CDI mode using 
Eq. (\ref{cons}), and then we can approximate for $k \ll \mathcal{H}$.  
Using the sum of the adiabatic and CDI modes for $V_\phi^{rad}$ in Ineq. (\ref{conditionscalar}), we 
find that the most negative $\mu$ can be is 
$ \sim -2 \times 10^{-5}$ (for the same parameters listed for the purely adiabatic case above), not much less compared to the exclusively 
adiabatic contribution of $ \sim- 3 \times 10^{-6}$.   

For the other isocurvature modes, the initial conditions for the perturbations have the same functional form since they are expressed as power series, but 
the coefficients of their terms (ignoring the constant of integration specified by the isocurvature primordial power spectrum for 
each mode) of the perturbations are considerably smaller in magnitude compared to those of the CDI mode, so their maximal contribution allowed by observations does 
not significantly alter the chances of satisfying Ineq. (\ref{conditionscalar}).

So we conclude that phantom dark energy with positive kinetic energy is not 
possible during the radiation era for constant $w$ with 1st-order corrections.  

\begin{center}
{\it Matter Domination}
\end{center}

During matter domination ($\mathcal{H} = 2/\tau$ and $\tau = (\frac{3 a}{\rho_{m0} \pi })^{1/2}$), the metric perturbation $h = W (k\tau)^2$ (leaving out the gauge modes) 
is obtained from 
Eqs. (\ref{perturb}) and (\ref{cons}).  
$V_\phi^{matt}$ during matter domination with constant $w_\phi \neq -1$ and sound 
speed $c_{s \phi}$ is obtained from Eq. (\ref{cons}) and $h$.  
We set the constants of integration for the decaying modes of the full solution for $V_\phi^{matt}$ to $0$, and this gives a real solution for $V_\phi^{matt}$.  For $k \ll \mathcal{H}$, $V_\phi^{matt}$ is 
\begin{equation}
\label{Vm2}
V_\phi^{matt} \approx \frac{3^{3/2} c_{s _\phi}^2~ W k^3 a^{3/2}}{2^{5/2} (\rho_{m0} \pi)^{5/2} ~(5+9c_{s _\phi}^2 - 15w_\phi)},
\end{equation}
where $W$ is a constant of integration (from $h$) and $\rho_{m0}= \frac{3 H_0}{8 \pi} \Omega_{m0}$ is the present-day density of matter.  We use $\Omega_{m0}=0.315$, 
which is consistent with Planck's reported value \cite{Planck}.  

Using Eq. (\ref{conditionscalar}), the condition for a positive kinetic term is 
\begin{equation}
\label{Vm2ineq}
\mu = - \frac{3 W (c_{s _\phi}^2 (1+6w_\phi)-5) k^2 a}{4 \pi \rho_{m0} (5+9 c_{s _\phi}^2 -15w_\phi)} \leq -1.
\end{equation}
For $w_\phi<-1$, we see that the inequality can only be satisfied for $W<0$.  Since $h$ goes like to a constant with respect to time multiplied by $(k \tau)^2$ for 
adiabatic perturbations for both the radiation and matter eras, we see that $W=A$ via continuity of $h$, and Eq. (\ref{IC}) also applies to $W$.  
For negative 
$W$, $\mu$ is
most negative for most negative $w_\phi$ (which would be $w_\phi \approx -2$ from observational constraints), largest $a$ 
(which is $a=a_{DE} = (\frac{-\Omega_{m0}}{\Omega_{DE0} 3 w_\phi})^{\frac{-1}{3w_\phi}}$, which is when the densities of matter and dark energy are equal and 
dark energy begins to dominate for constant $w_\phi$), and largest $k$ ($k \approx \mathcal{H}(a_{DE})$).  Using these parameters along with $W$ from Eq. (\ref{IC}), 
we find that the 
lefthand side of Eq. (\ref{Vm2ineq}) is $\sim 3 \times 10^{-6}$, so Ineq. (\ref{Vm2ineq}) is not satisfied.  Although our approximation $k \ll \mathcal{H}$ breaks down near the 
end of matter domination since $\mathcal{H}(a_{DE}) \sim 10^{-4}$, our conclusion still holds, and for smaller values of $a$ during matter domination 
for which the approximation is more accurate, Ineq. (\ref{Vm2ineq}) is certainly not satisfied.  

When including the maximal contribution from isocurvature (and we deal with the CDI mode for the reasons discussed in the previous section), we find $W$ from 
matching $V^{rad}_{\phi~ tot}(a_{eq},k)=V^{matt}_{\phi}(a_{eq},k)$, where $V^{rad}_{\phi~ tot}$ is the sum of the adiabatic and CDI contributions for $V^{rad}_\phi$, 
and we use the best fit that we used in the previous section on the radiation era.  
This matching barely changes the magnitude of $W$ from what it was in the purely adiabatic case, and $\mu$ changes from $\sim -3 \times 10^{-6}$ to 
$\sim -1 \times 10^{-5}$, leaving Ineq. (\ref{Vm2ineq}) still unsatisfied.  

Notice that the constant of integration $A$ from radiation domination and $W$ from matter domination 
must be oppositely signed in order for the kinetic energy to be positive during both periods.  And we found $W$ from the condition $W=A$ for adiabatic perturbations, 
and this is approximately true also when isocurvature contributions, which are small, are considered.  So we see that positive kinetic energy during radiation domination 
is incompatible with positive kinetic energy during matter domination for $k \ll \mathcal{H}$.  

So we conclude that positive kinetic energy is not 
possible during the matter era for constant $w_\phi$ with 1st-order corrections.

\begin{center}
{\it Dark Energy Domination}
\end{center}

During dark energy domination ($\mathcal{H} =  \frac{2}{(3w_\phi+1) \tau}$, $\tau= \frac{1}{1+3w_\phi} \sqrt{\frac{3}{2 \pi \rho_{DE0}}} a^{\frac{1+3w_\phi}{2}}$), 
$V_\phi^{DE}$ with constant $w_\phi \neq -1$ and sound speed $c_{s _\phi}$ is obtained from Eqs. (\ref{perturb1}), (\ref{perturb2}), (\ref{perturb5}), and (\ref{perturb6}): 
\begin{equation}
\label{VDEa}
V_\phi^{DE} = S \biggl(\frac{2\pi}{3} \rho_{DE0} \biggr)^{\frac{1}{3w_\phi +1}} a^{-1},
\end{equation}
where $\rho_{DE0} = \frac{3 H_0^2}{8 \pi} \Omega_{DE0}$ is the present-day density of dark energy.  We use 
$\Omega_{DE0} = 0.685$, which is consistent with Planck's reported values \cite{Planck}.  $V_\phi^{DE}$ is given without making the approximation 
$k \ll \mathcal{H}$.  It has two other modes that are purely imaginary, so we set 
their constants of integration to $0$ in order to have a real expression in Eq. (\ref{VDEa}).  

Inequality (\ref{conditionscalar}) becomes
\begin{equation}
\label{conditionconstw}
\mu = -S  \biggl(\frac{2 \pi}{3} \rho_{DE0} \biggr)^{\frac{1}{3w_\phi+1}} \frac{ (\frac{8 \pi}{3} \rho_{DE0})^{1/2}}{k}  (3+3w_\phi) a^{-\frac{3+3w_\phi}{2}}   \leq -1.
\end{equation}

The only way for this inequality to be satisfied is if $S <0$.  $S$ is found through matching $V_\phi^{matt}(a_{DE}) = V_\phi^{DE}(a_{DE})$ for 
$10^{-4}~\mathrm{Mpc}^{-1}~ \leq k \leq 8 \times 10^{-4}~\mathrm{Mpc}^{-1}$ and $-2 \leq w \leq -1$, our allowed ranges for $k$ and $w_\phi$.  We do the 
matching for the full solution for $V_\phi^{matt}$ since $k \ll \mathcal{H}$ is not accurate at $a_{DE}$.  

For either the purely 
adiabatic case or the mixed case with maximal isocurvature contribution, this matching implies $10^{-10} \lesssim S \lesssim 10^{-5}$, and this implies 
$- 6 \times 10^{-4} < \mu < - 5 \times 10^{-8}$, leaving Ineq. (\ref{conditionscalar}) unsatisfied.  

In fact, using Eq. (\ref{perturb6}), one finds that $\mu = \delta_\phi^{DE}$.  So $\mu \leq -1$ implies $\delta_\phi^{DE} \leq -1$, which would break our assumption 
of $|\delta_\phi^{DE}| \ll 1$ for perturbation theory, independent of $S$.  

So we conclude that positive kinetic energy for dark energy as a scalar field is not possible using 1st-order cosmological perturbation theory for the FLRW metric and
 constant $w_\phi$ and $c_{s _\phi}^2$.

\bigskip
\begin{center}
{\bf Non-Constant $w_\phi$ Models}
\end{center}

For models with $\dot{w}_\phi \neq 0$, the dynamics of Ineq. (\ref{conditionscalar}) are different due to the $\frac{dw_\phi}{da}$ term and $w_\phi$'s effect on $V_\phi$.  

Dark energy with constant $w_\phi<-1$ has density that behaves as a power law in $a$ as $\rho_\phi = \rho_{\phi 0}~ a^{-3(w_\phi+1)}$.  
All these models lead to a big rip \cite{bigrip} 
since $\rho_\phi \rightarrow \infty$ at a finite time in the future and all bound structures are ripped apart before that time due to the increasing repulsive force of dark energy.  
We consider in this section two models with $\rho_\phi$ that increase more slowly than any power law in $a$.  We examine a model leading 
to a little rip \cite{littlerip, littlerip2}, which means that $\rho_\phi \rightarrow \infty$ 
as $t \rightarrow \infty$ and all bound structures are ripped apart eventually.  We also consider a model leading to a pseudo-rip \cite{pseudorip}, which means that 
$\rho_\phi \rightarrow \rho_\infty$ as $t \rightarrow \infty$ and all bound structures at or below the threshold determined by $\rho_\infty$ rip apart.  

The little rip parametrization we consider is given by
\begin{equation}
\label{littleripeq}
\rho_{lr}=\rho_{DE0} \biggl(\frac{3 \alpha}{2 \rho_{DE0}^{1/2}} \ln a +1 \biggr)^2.
\end{equation}
This parametrization was fitted to recent supernovae data with the best fit of $\alpha = 2.26 \times 10^{-6} ~\mathrm{Mpc}^{-1}$ \cite{littlerip}.  The fitting essentially 
ensures $\dot{w}_{lr} \approx 0$ around $a=1$.  And note that $\rho_{lr}(a=1)=\rho_{DE0}$ by construction.  The equation-of-state parameter, obtained from Eqs. (\ref{Feq1}) 
and (\ref{Feq2}), is 
\begin{equation}
\label{lreos}
w_{lr} \equiv -1 - \frac{a \rho_{lr}'(a)}{3 \rho_{lr}} = -1 - \frac{1}{\frac{3}{2} \ln a + \sqrt{\rho_{DE0}}/\alpha}.
\end{equation}

The pseudo-rip parametrization we consider is given by
\begin{equation}
\label{pseudoripeq}
\rho_{pr} = \rho_{DE0} \biggl(\frac{\ln [\frac{1}{f+1/a} + \frac{1}{b}]}{\ln[ \frac{1}{f+1} + \frac{1}{b}]} \biggr)^s,
\end{equation}
where $f=10^{-23}$ and $s=48$.  This parametrization was also fitted to recent supernovae data with the best fit of $b= 0.01078$ \cite{pseudorip}, and 
$\rho_{pr}(a=1)=\rho_{DE0}$ by construction.  The equation-of-state parameter is 
\begin{equation}
\label{preos}
w_{pr} \equiv -1 - \frac{a \rho_{pr}'(a)}{3 \rho_{pr}} = -1- s b \frac{a}{3 (1+a f) (1+a (b+f)) \ln \biggl[\frac{1}{\frac{1}{a}+f} + \frac{1}{b}\biggr]}.
\end{equation}
Note that both $w_{lr}$ and $w_{pr}$ are strictly less than $-1$ for all $a$, and they approach $-1$ as $a \rightarrow \infty$.  

\begin{figure}
\begin{center}
\fbox{\includegraphics[scale=1.2]{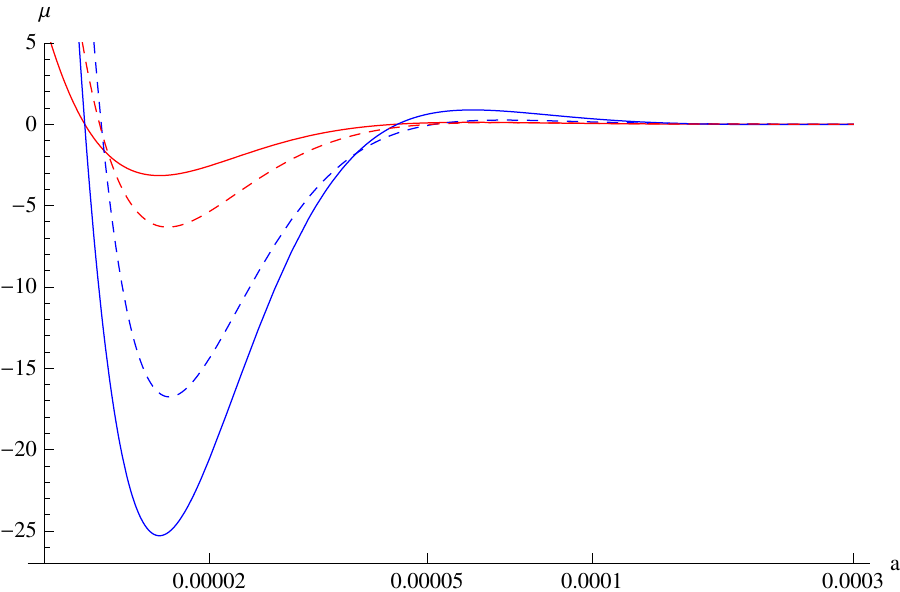}}
\caption{We plot $\mu$, the lefthand side of Ineq. (\ref{conditionscalar}), for $k=10^{-4}~ \mathrm{Mpc}^{-1}$ (blue lines) and $k=8 \times 10^{-4}~ \mathrm{Mpc}^{-1}$ 
(red lines) with chosen initial conditions $V_\phi^{rad}(10^{-5})=10^{-2}$ and 
$\dot{V}_\phi^{rad}(10^{-5})=2 \times 10^{-4}$ during the radiation era for adiabatic perturbations for both the 
little rip (solid lines) and pseudo-rip (dashed lines) parametrizations from Eqs. (\ref{lreos}) and (\ref{preos}).  All perturbations are sufficiently 
small.  The top set of patterned lines is red.  (Color plots available in the online 
version of this work.)  We see that the condition for positive kinetic energy can be satisfied for some but not all of the radiation era.}
\label{lrradfig}
\end{center}
\end{figure}

\begin{figure}
\begin{center}
\fbox{\includegraphics[scale=1.2]{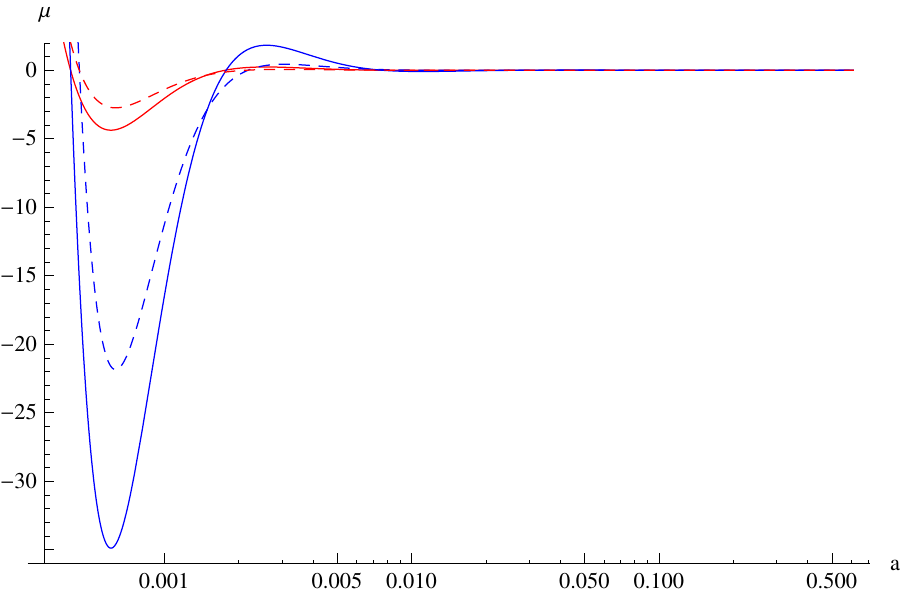}}
\caption{We plot $\mu$, the lefthand side of Ineq. (\ref{conditionscalar}), for $k=10^{-4}~ \mathrm{Mpc}^{-1}$ (blue lines) and $k=8 \times 10^{-4}~ \mathrm{Mpc}^{-1}$ 
(red lines) with chosen initial conditions $V_\phi^{matt}(a_{eq})=0.4$ and 
$\dot{V}_\phi^{matt}(a_{eq})=2 \times 10^{-3}$ during the matter era for adiabatic perturbations for both the 
little rip (solid lines) and pseudo-rip (dashed lines) parametrizations from Eqs. (\ref{lreos}) and (\ref{preos}).  All perturbations are sufficiently 
small.  The top set of patterned lines is red.  (Color plots available in the online 
version of this work.)  We see that the condition for positive kinetic energy can be satisfied for some but not all of the matter era.}
\label{lrmattfig}
\end{center}
\end{figure}

\begin{figure}
\begin{center}
\fbox{\includegraphics[scale=1.2]{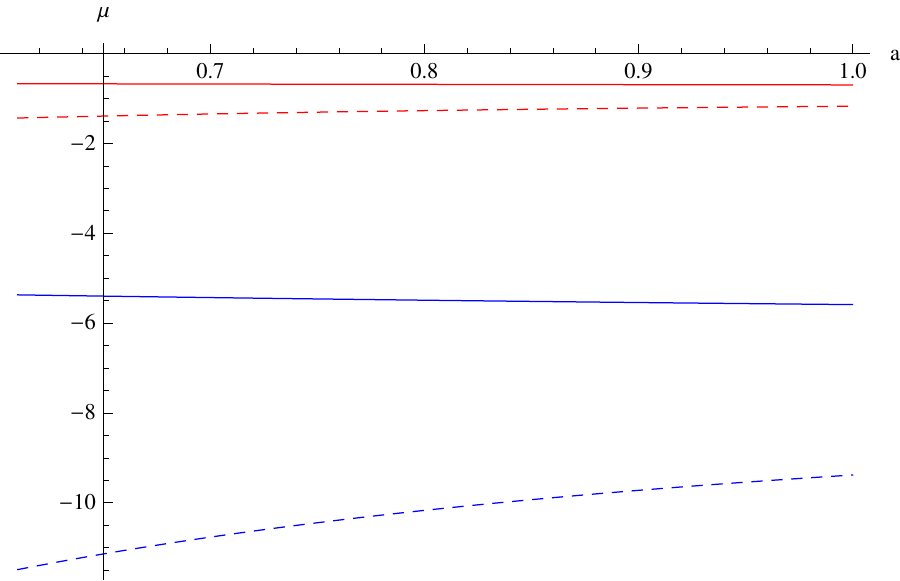}}
\caption{We plot $\mu$, the lefthand side of Ineq. (\ref{conditionscalar}), for $k=10^{-4}~ \mathrm{Mpc}^{-1}$ (blue lines) and $k=8 \times 10^{-4}~ \mathrm{Mpc}^{-1}$ 
(red lines).  We chose $V_\phi^{DE}(a_{DE})=0.6$, $\dot{V}_\phi^{DE}(a_{DE})= -3 \times 10^{-4}$, and $\ddot{V}_\phi^{DE}(a_{DE})= -8 \times 10^{-8}$ during the dark energy 
era for the 
little rip parametrization (solid lines) and $V_\phi^{DE}(a_{DE})=0.6$, $\dot{V}_\phi^{DE}(a_{DE})= -6 \times 10^{-4}$, and 
$\ddot{V}_\phi^{DE}(a_{DE})= -2 \times 10^{-7}$ for the pseudo-rip parametrization (dashed lines).  All perturbations are sufficiently 
small.  The top set of patterned lines is red.  (Color plots available in the online 
version of this work.)  We chose initial conditions such that the little rip parametrization does not have $\mu<-1$ for all $k$ while the pseudo-rip parametrization does.  
We see that the condition for positive kinetic energy can (apparently) be satisfied for all of the dark energy era, but see the text for further details.}
\label{lrdefig}
\end{center}
\end{figure}

We consider adiabatic perturbations during the radiation era ($h= A (k \tau)^2$), matter era ($h= W (k \tau)^2$), and dark energy era 
($h$ put in terms of $V_\phi$ and its derivatives using Eqs. (\ref{perturb1}), (\ref{perturb2}), (\ref{perturb5}), and (\ref{perturb6})) for the little rip and pseudo-rip parametrizations, 
and we numerically solve the differential 
equation we obtain for $V_\phi$.  When solving for a model with non-constant $w_\phi$, $V_\phi$ is no longer analytic in general, and one 
must specify initial conditions beyond the one parameter that is constrained by the primordial power spectrum in the radiation and matter eras.  
For reasonable initial conditions for both parametrizations
(given in the captions of Figs. (\ref{lrradfig}), (\ref{lrmattfig}), and (\ref{lrdefig})) and 
$A$ and $W$ constrained from the primordial power spectrum, we plot $\mu$ in Figs. (\ref{lrradfig}), (\ref{lrmattfig}), and (\ref{lrdefig}) for both $k=10^{-4}~\mathrm{Mpc}^{-1}$ 
and $k=8 \times 10^{-4}~\mathrm{Mpc}^{-1}$, the upper and lower bound of observationally relevant $k$ that we consider.  
For the radiation and matter eras, the inequality is satisfied for some of the era (where $\mu<-1$), but not for the whole era; this is the case for all $k$, 
independent of the initial conditions chosen 
for $V_\phi^{rad}$ and $V_\phi^{matt}$ and the signs of $A$ and $W$.  Adding in the isocurvature mode for $V_\phi^{rad}$ (which uses $h$ from Eq. (\ref{CDI})) and 
propagating this addition through the next eras does not change the magnitude 
or shape of the results in any of the three figures very much at all.  

\begin{figure}
\begin{center}
\fbox{\includegraphics[scale=1.2]{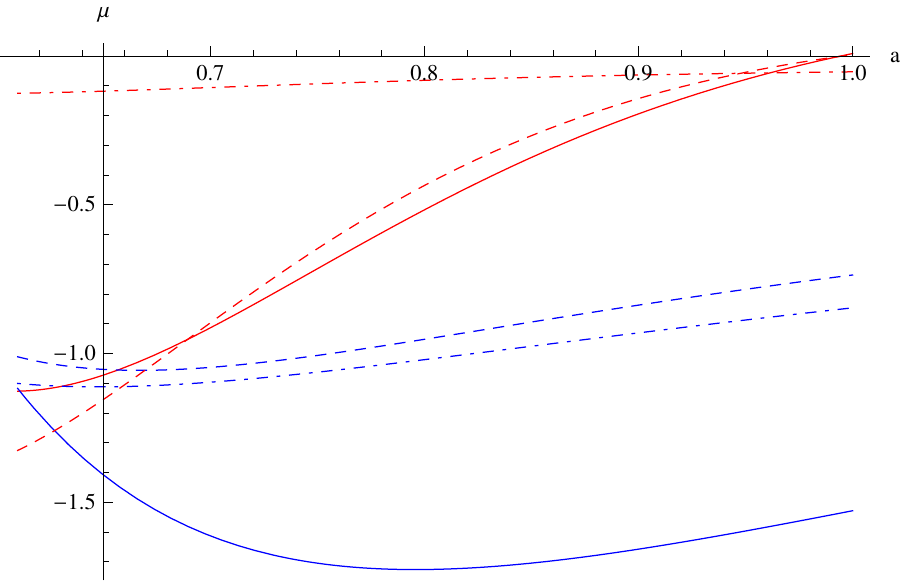}}
\caption{We plot $\mu$, the lefthand side of Ineq. (\ref{conditionscalar}), for $k=10^{-4}~ \mathrm{Mpc}^{-1}$ (blue lines) and $k=8 \times 10^{-4}~ \mathrm{Mpc}^{-1}$ 
(red lines) during the DM-DE era for the little rip parametrization (solid lines), pseudo-rip parametrization (dashed lines), and constant $w_\phi=-1.1$ (dot-dashed lines) 
for $a \in (0.61,1)$, where $0.61$ is close to $a_{DE}$ for all 3 parametrizations.  We chose the following initial conditions for 
$k=10^{-4}~ \mathrm{Mpc}^{-1}$:  little rip: $V_\phi^{DM-DE}(0.61)=0.24$, $\dot{V}_\phi^{DM-DE}(0.61)= -1.0 \times 10^{-4}$, $\ddot{V}_\phi^{DM-DE}(0.61)= -2.9 \times 10^{-4}$, 
$\dddot{V}_\phi^{DM-DE}(0.61)=5.3 \times 10^{-3}$; pseudo-rip:  $V_\phi^{DM-DE}(0.61)=0.22$, $\dot{V}_\phi^{DM-DE}(0.61)= -1.1 \times 10^{-4}$, 
$\ddot{V}_\phi^{DM-DE}(0.61)= 1.8 \times 10^{-4}$, $\dddot{V}_\phi^{DM-DE}(0.61)=1.3 \times 10^{-3}$; constant $w_\phi$:  
$V_\phi^{DM-DE}(0.61)= -7.3 \times 10^{-2}$, $\dot{V}_\phi^{DM-DE}(0.61)= -4.0 \times 10^{-5}$, $\ddot{V}_\phi^{DM-DE}(0.61)= 4.3 \times 10^{-5}$, 
$\dddot{V}_\phi^{DM-DE}(0.61)=9.7 \times 10^{-4}$.  We choose the following initial conditions for $k=8 \times 10^{-4}~ \mathrm{Mpc}^{-1}$:  
little rip: $V_\phi^{DM-DE}(0.61)=0.0$, $\dot{V}_\phi^{DM-DE}(0.61)= -4.5 \times 10^{-4}$, 
$\ddot{V}_\phi^{DM-DE}(0.61)= 8.0 \times 10^{-4}$, $\dddot{V}_\phi^{DM-DE}(0.61)=2.7 \times 10^{-2}$; pseudo-rip:  $V_\phi^{DM-DE}(0.61)=0.0$, 
$\dot{V}_\phi^{DM-DE}(0.61)= -5.3 \times 10^{-4}$, 
$\ddot{V}_\phi^{DM-DE}(0.61)= 2.7 \times 10^{-3}$, $\dddot{V}_\phi^{DM-DE}(0.61)=1.4 \times 10^{-2}$; constant $w_\phi$:  
$V_\phi^{DM-DE}(0.61)= - 7.2 \times 10^{-2}$, $\dot{V}_\phi^{DM-DE}(0.61)= -3.6 \times 10^{-5}$, $\ddot{V}_\phi^{DM-DE}(0.61)= 1.2 \times 10^{-4}$, 
$\dddot{V}_\phi^{DM-DE}(0.61)=9.4 \times 10^{-4}$.  All perturbations are sufficiently 
small.  The top set of patterned lines is red.  (Color plots available in the online 
version of this work.)  We see that the condition for positive kinetic energy can be satisfied for some but not all of the DM-DE era.}
\label{lrdmdefig}
\end{center}
\end{figure}

In Fig. (\ref{lrdefig}), we plot $\mu$ from the beginning of the dark energy domination era to $a=1$ (present day).  We find that one can choose a set of initial conditions such 
that $\mu<-1$ for all of $k \in(10^{-4}~\mathrm{Mpc}^{-1},~ 8 \times 10^{-4}~\mathrm{Mpc}^{-1})$ 
while still meeting the requirements of smallness for perturbation theory.  Even though $w_{lr}$ and $w_{pr}$ fail to provide positive kinetic energy for the whole of the 
radiation and matter eras, in theory, dark energy could be a phenomenon that is active only during its domination era and not before 
(as we have no observational evidence for dark energy's existence before its era, strictly speaking).  If dark energy becomes active only during dark energy domination as the 
result of, say, some spontaneous symmetry 
breaking, then there is no necessary continuity with the dark energy perturbations we calculated for the matter era.  Fig. 
(\ref{lrdefig}) indicates that it may be possible to have phantom dark energy that satisfies a field theory with positive kinetic energy for all relevant observational $k$ and $a$ (that 
is, up to the present).  However, 
in the future ($a>1$) for the models we examine, $\mu$ generally becomes $>-1$.  Also, one can be more precise by considering an era dominated by both matter and dark energy.  

\begin{center}
{\it DM-DE Domination Era}
\end{center}

We also consider an era with both dark matter and dark energy, $\mathcal{H} = a (\frac{8 \pi}{3} (\rho_{DE}(a)+ \rho_{c0}~ a^{-3}))^{1/2}$, 
where $\rho_{DE}(a)$ is the density of dark energy for a generic $w_\phi(a)$ (constant or not), and $\rho_{c0}= \frac{3 H_0^2}{8 \pi} \Omega_{c0}$ is the 
present-day density of dark matter.  We use $\Omega_{c0}=0.266$, which is consistent with Planck's reported values \cite{Planck}.  
Because $\theta_c=0$ in the synchronous gauge and $\delta P_c=0$, Eqs. (\ref{perturb}) and (\ref{cons}) are identical to 
the ones used for the dark energy era except for Eq. (\ref{perturb1}), which contains contributions from both 
$\delta_c$ and $\delta_\phi$ on the righthand side.  We use Eqs. (\ref{perturb1}), (\ref{perturb2}), (\ref{perturb5}), and (\ref{perturb6}) to obtain our differential equation in 
terms of $V_\phi$ and its derivatives.  

We have already seen that positivity of kinetic energy is not satisfied for constant $w_\phi$ during the matter and dark energy eras, so one would expect a similar outcome for the 
DM-DE era.  Indeed, one can match $V_\phi^{rad}(a_{eq})=V_\phi^{DM-DE}(a_{eq})$ along with the derivatives to set the initial conditions for solving the differential equation, 
and we find the $\mu$ is not $<-1$ by a wide margin.  

However, if dark energy is considered active only from $a_{DE}$ onward, then $V_\phi^{DE-DM}$ is not necessarily constrained from 
continuity with $V_\phi$ and its derivatives before this era, leaving free the choice of initial conditions.  Along with trying to achieve $\mu<-1$ for $a \in (a_{DE},1)$ and 
$k \in (10^{-4}~\mathrm{Mpc}^{-1},~  8 \times 10^{-4}~\mathrm{Mpc}^{-1})$, one has to satisfy the smallness of dark energy perturbations and the smallness of $\delta_c$, 
the latter of which can be checked 
via the use of Eq. (\ref{perturb1}).  After scanning the parameter space of initial conditions for $V_\phi^{DM-DE}$, 
we found that the positivity of kinetic energy is not satisfied for the relevant ranges of both $a$ and $k$, even if it can be satisfied for some $a$ and $k$.  And even 
for initial conditions that cause $\mu<-1$ for all $a \in (a_{DE},1)$ for a given $k$, $\mu$ still grows to be $>-1$ for $a>1$, and $|\delta_c|$ is often $>1$ for 
other values of $k$ with the same initial conditions.  In Fig. (\ref{lrdmdefig}), 
we plot $\mu$ for $w_{lr}$, $w_{pr}$, and constant $w_\phi$ for sample initial conditions.


\bigskip
\begin{center}
{\bf Dark Energy with $w_\phi>-1$}
\end{center}

\begin{figure}
\begin{center}
\fbox{\includegraphics[scale=1.2]{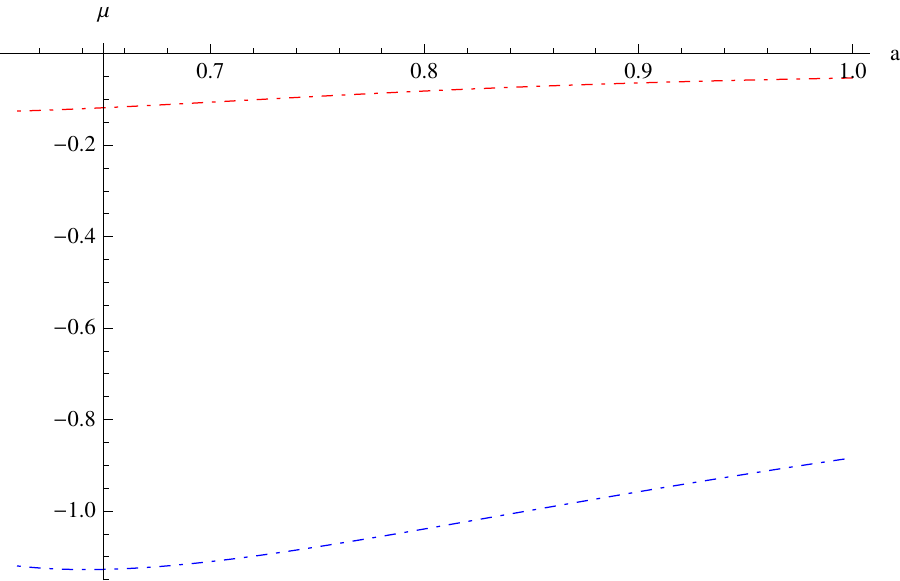}}
\caption{We plot $\mu$, the lefthand side of Ineq. (\ref{conditionscalar}), for $k=10^{-4}~ \mathrm{Mpc}^{-1}$ (blue line) and $k=8 \times 10^{-4}~ \mathrm{Mpc}^{-1}$ 
(red line) during the DM-DE era for constant $w_\phi=-0.99$ (dot-dashed lines) 
for $a \in (0.61,1)$ and for the same initial conditions for the constant $w_\phi$ cases in Fig. (\ref{lrdmdefig}). All perturbations are sufficiently 
small.  The top patterned line is red.  (Color plots available in the online 
version of this work.)  We see that positive kinetic energy is not satisfied for all of the DM-DE era.}
\label{quintdmdefig}
\end{center}
\end{figure}

It is interesting to note that for the case of $w_\phi>-1$, the only change in Ineqs. (\ref{condition3}) and (\ref{conditionscalar}) is $\leq~ \rightarrow~ \geq$, 
and the magnitudes 
of $\mu$ of our best attempts to satisfy positivity of kinetic energy for constant $w_\phi$ does not increase 
substantially with a slight change in the magnitude of $w_\phi$ from a value less than $-1$ to one bigger than $-1$.  Since $V_\phi$ was computed in different eras for 
a generic $w_\phi \neq -1$, they apply for both $w_\phi<-1$ and $w_\phi>-1$.  So for $w_\phi>-1$, it is easy to see that 
$\mu \geq -1$ for either choice of sign for the constants of integration during all single-component eras, meaning that 
negative kinetic energy is avoided for constant $w_\phi >-1$.  However, for constant $w_\phi>-1$ in the DM-DE era, we have checked that $\mu$ 
can be $< -1$ (negative kinetic energy for $w_\phi>-1$) for certain scales and times with the right choice of initial conditions while the smallness of perturbation theory is satisfied, just as we analogously found 
$\mu<-1$ (positive kinetic energy for $w_\phi<-1$) possible in this era for constant $w_\phi<-1$ for certain scales and times!  See Fig. (\ref{quintdmdefig}) for an example 
of negative kinetic energy from a $w_\phi>-1$ model.  Note that the constant $w_\phi$ lines from Figs. (\ref{lrdmdefig}) and (\ref{quintdmdefig}) are very similar.  
We suspect that 
non-constant parametrizations for non-phantom dark energy can also give rise to particular scales and times during which kinetic energy is negative.  
This 
prospect of quintessence dark energy's violation of the positivity of kinetic energy is intriguing and should be pursued in future research\footnote{For the case of $w_\phi=-1$, 
$\delta P_\phi/\rho_\phi = c_{s \phi}^2 \delta_\phi$, 
and Ineq. (\ref{condition}) becomes $1 + \delta_\phi(1+ c_{s \phi}^2) + w_\phi \geq 0$.  If we multiply every term in Eq. (\ref{perturb6}) by $(1+w)$, 
we see that $\delta_\phi = 0$ for $\sigma_\phi=0$, and Ineq. (\ref{condition}) becomes $1+w_\phi = 0 \geq 0$.  So negative kinetic energy is avoided for $w_\phi = -1$.}.  

\bigskip
\begin{center}
{\bf Conclusion}
\end{center}

In this work, we have studied models of phantom dark energy completely within the confines of general relativity to discern whether or not they can be represented 
as a scalar field theory with positive kinetic energy.  While it is clear for FLRW space at 0th order that every phantom model of dark energy fails to give a positive kinetic 
term for a scalar field theory, it is not as clear at 1st order.  For a scalar field defined for phantom dark energy for the more observationally accurate 
1st-order FLRW space and {\it not} for 0th order, it is mathematically possible to satisfy positivity of kinetic energy ($\mu<-1$) for certain relevant ranges of $a$ and $k$.  

For constant $w_\phi$ models during the radiation and matter eras constrained by the primordial power spectrum, positivity of kinetic energy is not satisfied, and for 
the dark energy era, perturbation theory would have to be violated in order for $\mu <-1$ since $\mu=\delta_\phi^{DE}$.  For the DM-DE era, $\mu>-1$ if $V_\phi^{DM-DE}$ is matched up with the observationally constrained $V_\phi^{rad}$ for models with constant 
$w_\phi$, and positivity of kinetic energy is not satisfied.

For the models with non-constant $w_\phi$ we consider during the radiation and matter eras, despite the freedom to choose initial conditions while still 
meeting constraints from the primordial 
power spectrum, we find that positivity of kinetic energy is satisfied only for limited 
ranges of $a$ and $k$.  During the dark energy era, our models with non-constant $w_\phi$ can satisfy positivity of kinetic energy for all relevant $a$ and observational $k$ 
given the right initial conditions.  
But generally 
$\mu>-1$ for the future ($a>1$), and the more accurate description of the DM-DE era provides different results.  For our models with constant or non-constant 
$w_\phi$ during the DM-DE era, positivity of kinetic energy can be satisfied only for some $a$ and $k$ without violating perturbation theory's 
assumption of smallness, and $\mu>-1$ in the future.  In fact, one could be even more accurate and assume contributions to Eq. (\ref{perturb}) and $\mathcal{H}$ 
from radiation, matter, and dark energy, and the added constraints on the dark energy perturbations from the requirement of smallness for radiation and matter perturbations may 
make the satisfaction of the positivity of kinetic energy even more difficult.  

For the case of $w_\phi>-1$, we saw that violation of the positivity of kinetic energy is possible and should be further examined.  We 
suspect that the more accurate assumption of contributions to Eqs. (\ref{perturb}) and $\mathcal{H}$ from radiation, matter, and dark energy and constraints 
from the smallness of all the components' perturbations would make the satisfaction of negative kinetic energy for non-phantom parametrizations more difficult.     

So overall, in the context of general relativity, we find that phantom dark energy does not provide a scalar field theory with positive kinetic energy in 
1st-order FLRW space for all relevant scales and times, at least for constant-$w$ models and the others we have tested, but 
we do point out that phantom dark energy is not categorically indicative of negative kinetic energy for a scalar field theory, and, as far as we know, 
this work is the first attempt in explicitly checking the possibility of saving phantom dark energy as a scalar field theory purely through perturbations.   
By the same token, we found that quintessence models may have negative kinetic energy for certain times and scales, and this should be further 
explored in future work.  Other scenarios in the framework of 
FLRW cosmology can be studied rigorously for the positivity of the kinetic energy of dark energy which may serve to more precisely quantify the effects we demonstrate.  
Higher-order perturbations and the back-reaction of perturbations on the background would be an important consideration.  Additionally, if dark energy is not a perfect fluid, 
different conditions may apply for the positivity of its kinetic energy.  
And if dark energy and dark matter are coupled, the condition for positive kinetic energy changes, although 
some models may suffer from instabilities during the early universe, as \cite{0804.0232} points out.  For the covariant model they propose, perhaps the divergence in the 
gauge-invariant curvature perturbation can be pushed back early enough to a time for which quantum gravity dominates the relevant length scales so 
that there is no divergence for the classical regime.  Other frameworks (such as modified gravity, the presence of quantum effects, vector field dark energy, non-perturbative 
effects) and metrics 
(such as Bianchi and Tolman-Bondi models) for a universe deviating from a perfectly isotropic and homogeneous one can be studied 
to determine the viability of phantom dark energy.  Along with these different frameworks come different manifestations of observational constraints.  
We leave these other avenues to future work.  

\bigskip

\begin{center}
{\bf Acknowledgements}
\end{center}

KJL is supported by the Pirrung Foundation.

\end{document}